\documentclass[prd,showpacs,nofootinbib,preprintnumbers,preprint]{revtex4}

\def\be{\begin{equation}}
\def\ee{\end{equation}}
\def\ba{\begin{eqnarray}}
\def\ea{\end{eqnarray}}

\def\nn{\nonumber}
\def\lb{\label}

\def\nn{\nonumber}
\def\r{\hat{r}}

\def\C{{\cal C}}

\def\e{{\rm e}}
\def\x{{\dot x}^\mu}
\begin{document}
\title{\begin{flushright}\begin{small}    LAPTH-046/15
\end{small} \end{flushright} \vspace{1.5cm}
Rehabilitating space-times with NUTs}
\author{G\'erard Cl\'ement} \email{gerard.clement@lapth.cnrs.fr }
\affiliation{LAPTh, Universit\'e Savoie Mont Blanc, CNRS, 9 chemin de Bellevue, \\
BP 110, F-74941 Annecy-le-Vieux cedex, France}
\author{Dmitri Gal'tsov} \email{galtsov@phys.msu.ru}
\affiliation{Department of Theoretical Physics, Faculty of Physics,
Moscow State University, 119899, Moscow, Russia}
\author{Mourad Guenouche} \email{guenouche_mourad@umc.edu.dz }
\affiliation{ Laboratoire de Physique Th\'eorique, D\'epartement de
Physique, \\ Facult\'e des Sciences Exactes,
Universit\'e de Constantine 1, Algeria;  \\
Department of Physics, Faculty of Sciences, Hassiba Benbouali
University of Chlef, Algeria}

\begin{abstract}
We revisit the Taub-NUT solution of the Einstein equations without
time periodicity condition, showing that the Misner string is still
fully transparent for geodesics. In this case, analytic continuation
can be carried out through both horizons leading to a Hausdorff
spacetime without a central singularity, and thus geodesically
complete. Furthermore, we show that, in spite of the presence of a
region containing closed time-like curves, there are no closed
causal {\em geodesics}. Thus, some longstanding obstructions to
accept the Taub-NUT solution as physically relevant are removed.
\end{abstract}
\maketitle

\setcounter{equation}{0}
\section{Introduction}
The Taub-NUT solution of the vacuum Einstein equations with
Lorentzian signature \cite{Taub,NUT,misner,hawking} remains one of
the most puzzling results of General Relativity. It realizes the
idea of gravitational electric-magnetic duality  and is often
interpreted, by analogy with the Dirac magnetic monopole, as the
field of a gravitational dyon with the usual mass $m$ and a
``magnetic'' mass $n$ \cite{Dowker}. Astrophysicists gave tribute to
this solution suggesting to explore its signatures in the
microlensing data \cite{micro}. At the same time, most theorists
consider it as unphysical because of the presence of a Misner string
singularity on the polar axis (the gravitational analogue of the
Dirac string for the magnetic monopole), with still debated
features, and regions containing closed timelike curves (CTCs).

To make the string unobservable, Misner suggested to impose
periodicity on the time coordinate \cite{misner} which entails,
however, further serious problems. First, the space-time then
contains closed time-like curves everywhere. Second, with this
periodicity condition the analytically extended Taub-NUT spacetime
is either geodesically incomplete \cite{misner,hawking} (extension
can be carried out through only one of the two horizons), or can be
maximally extended to a geodesically complete but non-Hausdorff
spacetime \cite{geroch,hajicek}.

Another option is to preserve causality in the large by abandoning
the time periodicity condition, thereby retaining the Misner string
as an unremovable singularity. It was suggested by Bonnor
and others \cite{bonnor}, that the Misner string should be
interpreted as a singular material source of angular momentum.  On
the other hand, Miller et al. \cite{miller} have shown that the
vacuum Taub-NUT spacetime (without the time periodicity condition)
can be maximally extended \`a la Kruskal through both horizons.
Because the extended spacetime presents a coordinate singularity on
the polar axis, they considered it to be geodesically incomplete
\cite{miller,kagra}.

In this Letter, we consider motion in the Taub-NUT space without
time periodicity in greater detail and show that the Misner string
is fully transparent for geodesics hitting it. Since with this
interpretation the analytical continuation at the horizon is not
problematic and there is no central singularity, the whole extended
Taub-NUT space-time turns out to be {\em geodesically complete},
removing the major obstruction to give it physical significance.

The other problem with the Taub-NUT solution consists in the
presence of a region surrounding the Misner string containing CTCs,
which are generally considered to violate causality \cite{hawking}.
We show that for certain values of the parameter $C$ which was
previously used to fix the location of the Misner string, these CTCs
are not {\em causal geodesics}, and thus do not lead to causality
violations for a freely falling observer.

\section{The setup}
We start with the family of Taub-NUT spacetimes \be\lb{metric} ds^2
= - f(dt-2n(\cos\theta+C)\,d\varphi)^2 + f^{-1}dr^2 +
\rho^2(d\theta^2 + \sin^2\theta d\varphi^2)\,,\qquad \rho^2=r^2+n^2
 \ee
with $ f = (r^2-2rm-n^2)/\rho^2\,, $ where $m$ is the ordinary mass,
$n$ is the "magnetic" mass or NUT parameter and $C$ is an additional
parameter related to the ``large'' coordinate transformation $t\to
t+C\varphi$. Note that $C$ should be considered
as physical rather than pure gauge parameter, since it changes the
asymptotic behavior of the metric.
Its introduction was often used to modify
the position of the Misner string: for $C=-1$ it lies at the
southern hemisphere, for $C=1$ --- at the northern, for $C=0$ at
both of them.   Note also that the areal radius $(r^2+n^2)^{1/2}$ is
always finite, so the space-time has no central singularity.  This
is possible because the maximally analytically extended Taub-NUT
spacetime \cite{miller} has two distinct regions at spacelike
infinity $r \to \pm\infty$.

 The metric is symmetric \cite{misner,Dowker,Perry,ZS} under time
translations, generated by the Killing vector $K_{(t)}=\partial_t$,
and $so(3)$ local rotations associated with $K_{(a)},\,a=x,y,z,$
which can be compactly presented as \ba
K_{(\pm)}&=&K_{(x)}\pm i K_{(y)}=\e^{\pm i\varphi}\left( \pm i\partial_\theta -\cot\theta\,\partial_\varphi- \frac{2n(1+C \cos\theta)}{\sin\theta}\,\partial_t\right)\,,\nn\\
K_{(z)}&=&\partial_\varphi+2nC\,\partial_t\,, \ea The associated
four first integrals of geodesic equations $K_{(a)\mu}\x$ ($\x =
dx^\mu/d\tau$) read: \ba
E&=&\left({\dot t}-2n(\cos\theta +C){\dot{\varphi}}\right)f\,, \lb{E}\\
J_{\pm}&=&J_x\pm i J_y=\left(2nE\sin\theta -\rho^2(i{\dot\theta}-\sin\theta\cos\theta {\dot\varphi}) \right)\e^{\pm i\varphi}\,,\\
J_z&=&2nE\cos\theta+\rho^2\sin^2\theta
{\dot\varphi}\,,\lb{varphi} \ea with $J_x,\,J_y,\,J_z$ forming a
Cartesian vector $\vec{J}$ in isospace. This can be decomposed into the mutually
orthogonal orbital and ``spin'' parts \cite{ZS}
 \be\lb{J}
\vec{L} + \vec{S} = \vec{J}\,,\qquad \vec{L} =
\rho^2\,\r\wedge\dot{\r}\,, \qquad \vec{S} = 2nE\hat{r}\,,
 \ee
where $\r =
(\sin\theta\cos\varphi,\,\sin\theta\sin\varphi,\,\cos\theta)$ is a
unit vector normal to the two-sphere. It follows from the orthogonality of $\vec{L}$ and $\vec{S}$ that
 \be\lb{para}
\vec{J}\cdot\hat{r}=2nE\,.
 \ee
In the magnetic monopole case, such a first integral means that the
the trajectory of the charged particle lies on the surface of a cone
with axis $\vec{J}$ originating from the magnetic monopole source
$r=0$. However the Taub-NUT gravitational field has  no apex for the
``cone''. Rather this means that the geodesic intersects all the
two-spheres of radius $r$ on the same small circle, or parallel,
$\C$ with polar axis $\vec{J}$. Squaring (\ref{J}) leads to
 \be\lb{J2LE}
\vec{J}^2 = \vec{L}^2 + 4n^2E^2\,,
 \ee
which can be rewritten as
 \be\lb{l2}
\rho^4[\dot\theta^2 + \sin^2\theta\,\dot\varphi^2] = l^2\,,
 \ee
with $l^2 = J^2-4n^2E^2$ (denoting $J^2 = \vec{J}^2$). Inserting this into
the normalization condition $\x{\dot x}^\nu g_{\mu\nu}=\varepsilon$
($=-1$ for timelike and $0$ for null geodesics)
 leads to the effective radial equation
 \be\lb{rad1}
\dot{r}^2 + f(r)\left[\frac{l^2}{\rho^2}-\varepsilon\right] =
E^2\,,
 \ee
which is identical to the equation for radial motion in the
equatorial plane for the metric (\ref{metric}) without the term
$-2n\cos\theta d\varphi$.

\section{Misner string crossing}
Passing to the new parameter $\lambda$ on the geodesic defined by
$d\tau =\rho^2 d\lambda$, and putting $ \xi =\cos
\theta$ we obtain
\begin{equation}\lb{theta}
\left(\frac{d\xi}{d\lambda}\right)^{2} = -J^{2}\,\xi^2 +
4nEJ_{z}\,\xi + (l^{2}-J_{z}^{2})\,.
\end{equation}

Assuming $J^2\neq0$ ($J^2=0$ implies from (\ref{J2LE}) $E=0$ and
$l=0$), Eq. (\ref{theta}) is solved (up to an additive constant to
$\lambda$) by \cite{kagra}
 \be\lb{theta1}
\cos\theta = J^{-2}\left[2nEJ_z + lJ_{\bot}\cos(J\lambda)\right] = \cos\psi\cos\eta + \sin\psi\sin\eta\cos(J\lambda)\,,
 \ee
where
$
J_{\bot}^2 = J^2 - J_{z}^2\,, \tan\eta=l/2nE,\; \tan\psi=J_\bot/J_z.
$
Eq. (\ref{theta}) has two turning points $\theta_{\pm}$ such that
 \be\lb{thetapm}
\cos\theta_{\pm} = J^{-2}\left(2nEJ_z \pm lJ_{\bot}\right)=
\cos(\psi\mp\eta)\,.
 \ee
It follows that the trajectory crosses periodically the Misner
string, $\cos\theta_{\pm} = \pm1$ only if
 \be\lb{cross}
J_z = 2nE\,, \quad {\rm or}\quad  J_z = -2nE\,.
 \ee
The only geodesics which can cross both components of the Misner
string are those with $\eta=\pi/2$ ($E=J_z=0$), leading to
$\dot{t}=\dot{\varphi}=0\,$; according to (\ref{rad1}), in the
stationary sector ($f(r)>0$) these can only be spacelike geodesics.
The trajectory can also stay on the Misner string component
$\theta=0$ or $\pi$ if (\ref{cross})) is satisfied with $2nE=\pm J$.

The differential equation (\ref{varphi}) for $\varphi$ can be
rewritten as
 \be\lb{varphi1}
\frac{d\varphi}{d\lambda}
=\frac12\left[\frac{J_z-2nE}{1-\cos\theta(\lambda)} +
\frac{J_z+2nE}{1+\cos\theta(\lambda)}\right]\,,
 \ee
with $\cos\theta(\lambda)$ given by (\ref{theta1}). This is solved
by \cite{kagra}
 \be\lb{solvarphi}
\varphi - \varphi_0 =
\arctan\left[\frac{\cos\psi-\cos\eta}{1-\cos(\psi-\eta)}
\tan\,\frac{J\lambda}2\right]
+ \arctan\left[\frac{\cos\psi+\cos\eta}{1+\cos(\psi-\eta)}
\tan\,\frac{J\lambda}2\right]\,.
 \ee

For trajectories crossing the North Misner string, with  $J_z =
2nE$, this reduces to
 \be\lb{solvarphi2}
\varphi - \varphi_1 =
\arctan\left(\cos\eta\,\tan\left(\frac{J\lambda}2\right)\right)\,,
 \ee
with $\eta=\arcsin l/J,\;\varphi_1 = \varphi_0 -
{\rm{sgn}}(\tan(J\lambda/2))\pi/2$. A similar formula applies in the
case of the South Misner string, with $\eta$ replaced by $\pi-\eta$
and $J\lambda$ replaced by $J\lambda-\pi$ (note that according to
(\ref{theta1}) the North Misner string is crossed for
$\lambda=2k\pi/J$, while the South Misner string is crossed for
$\lambda=(2k+1)\pi/J$, $k$ integer). In the case e.g. of the North
Misner string, this gives on account of (\ref{theta1}),
 \be\lb{solvarphi3}
\cos(\varphi - \varphi_1) =
\frac{J_z}{J_{\bot}}\tan\left(\frac{\theta}2\right)\,,
 \ee
consistent with (\ref{para}) (the choice $\varphi_1=0$ in
(\ref{solvarphi3}) corresponds to the choice $\vec{J} =
(J_{\bot},0,J_z)$ in (\ref{para})).

When the parameter $\lambda$ varies over a period, e.g. $\lambda \in
[-\pi/J,\,\pi/J]$, the argument of the first or second $\arctan$ in
(\ref{solvarphi}) varies from $-\infty$ to $+\infty$ for
$J_z\mp2nE>0$, and from $+\infty$ to $-\infty$ for $J_z\mp2nE<0$. It
is identically zero for $J_z\mp2nE=0$. Accordingly, the variation of
$\varphi$ over a period is
 \be\lb{Dvarphi}
\Delta\varphi = \pi \left[{\rm sgn}(J_z-2nE) + {\rm sgn}(J_z+2nE)
\right]\,,
 \ee
leading to the possible values $|\Delta\varphi| = 2\pi$, $\pi$, or $0$.
These correspond precisely to the variations of the azimutal angle
as the test particle describes on the two-sphere a small circle which
may either circle the North or South pole, go straight though one
of these poles, or do not circle or cross the poles.
For $J_z^2>4n^2E^2$ ($|\Delta\varphi|=2\pi$) the
parallel $\C$ circles the North-South polar axis, i.e. the Misner
string. For $J_z^2<4n^2E^2$, ($|\Delta\varphi|=0$) $\C$ does not
circle the Misner string. And for $J_z=\pm2nE$
($|\Delta\varphi|=\pi$), $\C$ goes through the North or South pole,
as discussed above. Clearly the Misner string is
completely transparent to the geodesic motion!

 \section{Absence of closed causal geodesics}
 The ADM form of the metric (\ref{metric}) is
 \be\lb{adm}
ds^2 = - \frac{f\rho^2\sin^2\theta}{\Sigma}\,dt^2 + f^{-1}dr^2 +
\rho^2d\theta^2 +
\Sigma\left(d\varphi+\frac{2nf(\cos\theta+C)}{\Sigma}\,dt\right)^2,
 \ee
with
 $
\Sigma(r,\theta) = \rho^2\sin^2\theta - 4n^2f(\cos\theta+C)^2\,. $
For $f(r)<0$, $\Sigma$ is positive definite, while for $f(r)>0$
(outside the horizon), which which assume further, $\Sigma$ becomes
negative, and closed timelike curves (CTCs) appear, in a
neighborhood of the Misner string given by $\Sigma(r,\theta)<0$. The
surface $\Sigma(r,\theta)=0$ bounding this CTC neighborhood is a
causal singularity of the spacetime, where the signature of the
spacetime changes from $(-+++)$ outside to $(+++-)$ inside. This
singularity is, just as the Misner string itself, completely
transparent to geodesic motion. Nevertheless, the occurrence of CTCs
in a spacetime is usually considered to violate causality
\cite{hawking}. An observer travelling around such a CTC would
eventually return to his original spacetime position after a finite
proper time lapse, thus opening the possibility for time travel.
However, unless this observer is freely falling, such a CTC travel
would necessarily involve accelerations generated e.g. by rocket
engines. One can argue that the back-reaction of these matter
accelerations on the spacetime geometry would deform it in such a
way that chronology would ultimately be preserved. If this reasoning
is correct, causality violation can only occur in spacetimes with
closed timelike geodesics (CTGs), or possibly closed null geodesics
(CNGs). We now show that there are no closed timelike or null
geodesics in the Taub-NUT spacetime with $|C|\le1$.

Combining the Eqs. (\ref{E},\ref{varphi}) and passing to
$\lambda$-parametrization one is led to split $t(\lambda) =
t_r(\lambda) + t_{\theta}(\lambda)$ satisfying
 \ba
\frac{dt_{\theta}}{d\lambda} &=& 4n^2E +
n\left[\frac{(C+1)(J_z-2nE)}{1-\cos\theta(\lambda)} +
\frac{(C-1)(J_z+2nE)}{1+\cos\theta(\lambda)}\right]\,,\lb{ta1}\\
\frac{dt_r}{d\lambda} &=& E\frac{\rho^2}{f(r)}\,, \lb{tr} \ea with
$\cos\theta(\lambda)$ given by (\ref{theta1}). The explicit solution
to equation (\ref{ta1}) is \cite{kagra}
 \ba\lb{solta1}
t_{\theta}(\lambda) &=& 4n^2E\lambda +
2n(C+1) \arctan\left[\frac{\cos\psi-\cos\eta}{1-\cos(\psi-\eta)}
\tan\,\frac{J\lambda}2\right] \nn\\
&+& 2n(C-1)\arctan\left[\frac{\cos\psi+\cos\eta}{1+\cos(\psi-\eta)}
\tan\,\frac{J\lambda}2\right]\,,
 \ea
in the interval $-\pi/J < \lambda < \pi/J$. The resulting variation
of $t_{\theta}$ over a period $2\pi/J$ of $\lambda$ is
 \be\lb{Dttheta}
\Delta t_{\theta} = 2\pi n\left[\frac{4nE}J  + (C+1){\rm
sgn}(J_z-2nE) + (C-1){\rm sgn}(J_z+2nE) \right]\,.
 \ee
During a period $2\pi/J$ of the angular motion,
 \be\label{Dtang}
\Delta t_{\theta} \ge 4\pi n\left(\frac{2nE}J-1\right)
 \ee
for $|C|\le1$. Also, from (\ref{tr}) and (\ref{rad1}),
 \be
\frac{dt_r}{d\lambda} = \frac{E\rho^2}{f(r)} \ge E^{-1}[l^2 -
\varepsilon\rho^2] \ge E^{-1}[l^2 - \varepsilon n^2]
 \ee
in the stationary sector $f(r) > 0$. This leads to
 \be
\Delta t_r \ge \frac{2\pi}{EJ}\left[l^2 - \varepsilon n^2\right]\,,
 \ee
over the same period. Adding the two together, we obtain
 \be
\Delta t = \Delta t_r + \Delta t_{\theta} \ge \frac{2\pi}E\left[J -
2nE - \varepsilon n^2/J\right]\,.
 \ee
For $\varepsilon = -1$ this is clearly positive definite. For
$\varepsilon = 0$, this can vanish only for $2nE = J$ ($l=0$). But
in this case $\Delta t_{\theta} \ge 0$, while $dt_r/d\lambda$, and
thus also $\Delta t_r$, is positive definite. Thus, for $|C|\le1$
all timelike or null geodesics which stay in the stationary sectors
$r > r_h$ are causal (future directed).

The above reasoning fails for $|C|>1$, in which case the lower bound
(\ref{Dtang}) is replaced by $\Delta t_{\theta} \ge 4\pi
n(2nE/J-|C|)$. One can show \cite{wnut} that, for any parameter set
$(m,n)$, one can find a value of $C$ such that there are CNGs (and,
presumably, CTGs for larger values of $|C|$). For instance, for
$m=0$ and $C = -\sqrt3$, the circle $t={\rm const.}$, $r=\sqrt3n$,
$\theta=\arctan\sqrt2$ is a null geodesic.

\section{Conclusion}
We have shown that, contrary to longstanding prejudice (for a recent
discussion see \cite{kagra}), the Taub-NUT space-time without
periodic identification of time is geodesically complete. This is
valid for the whole family of metrics with arbitrary $C$, and
presumably can be extended to other spacetimes with NUT parameter.
Our proof is based on purely geometric properties of the solution
and is independent on whether or not one associates with the Misner
string some extra matter source. We also show that, in spite of the
presence of a region where the azimuthal coordinate is timelike and
the temporal coordinate spacelike, there are for $|C|\le1$ no closed
timelike or null {\em geodesics} which could violate causality.

We realize that our results do not remove all objections against
physical attribution of the metric with NUTs to the real world, in
particular, we do not consider quantum effects
\cite{Dowker,Perry}. Still, we hope that our findings remove some important
obstructions to recognition of these spacetimes as physically
relevant and will stimulate further work in this direction.

\section*{Acknowledgments} DG and MG would like
to thank LAPTh Annecy-le-Vieux for hospitality at different stages
of this work. DG also acknowledges the support of the Russian
Foundation of Fundamental Research under the project 14-02-01092-a.
MG acknowledges the support  of the  Ministry of Higher Education and
Scientific Research of Algeria (MESRS) under grant 00920090096.

\end{document}